\documentclass[10pt, english]{article}
\usepackage[margin=0.7in]{geometry}

\usepackage{epsfig}
\usepackage{times}
\usepackage{amsmath}
\usepackage{amssymb}
\usepackage{xspace}
\usepackage{algorithmicx}
\usepackage{algpseudocode}
\usepackage{subfigure}
\usepackage{graphicx}
\usepackage{caption}

\usepackage[latin1]{inputenc}
\usepackage{tikz}
\usetikzlibrary{shapes,arrows}

\newcommand {\mm}[1] {\ifmmode{#1}\else{\mbox{\(#1\)}}\fi}

\newsavebox{\smallProofsym}                            
\savebox{\smallProofsym}

\makeatletter
\long\def\@makecaption#1#2{%
  \vskip\abovecaptionskip
  \sbox\@tempboxa{\small #1: #2}%
  \ifdim \wd\@tempboxa >\hsize
    \small #1: #2\par
  \else
    \global \@minipagefalse
    \hb@xt@\hsize{\hfil\box\@tempboxa\hfil}%
  \fi
  \vskip\belowcaptionskip}
\makeatother

\newcommand{\R}        {\mm{{\mathbb R}}}

\newcommand{\X}        {\mm{{\mathbb X}}}

\newcommand{\Y}        {\mm{{\mathbb Y}}}

\newcommand{\Tcal}        {\mm{{\mathcal T}}}
\newcommand{\Vcal}        {\mm{{\mathcal V}}}
\newcommand{\Ecal}        {\mm{{\mathcal E}}}
\newcommand{\Fcal}        {\mm{{\mathcal F}}}
\newcommand{\Wcal}        {\mm{{\mathcal W}}}
\newcommand{\Mcal}        {\mm{{\mathcal M}}}
\newcommand{\Ncal}        {\mm{{\mathcal N}}}
\newcommand{\Ccal}        {\mm{{\mathcal C}}}

\newtheorem{definition}{Definition}[section]

\author{Paul Bendich\thanks{Department of Mathematics, Duke University, and Geometric Data Analytics, Inc.},
        Ellen Gasparovic\thanks{Department of Mathematics, Union College},
       Christopher J. Tralie\thanks{Department of Electrical and Computer Engineering, Duke University},
       and John Harer\thanks{Departments of Mathematics and of Electrical and Computer Engineering, Duke University, and Geometric Data Analytics, Inc.}
}

\title{Scaffoldings and Spines: Organizing High-Dimensional Data Using Cover Trees, Local Principal Component Analysis, and Persistent Homology}

\begin{document}

\maketitle

\begin{abstract}

We propose a flexible and multi-scale method for organizing, visualizing, and understanding datasets sampled from or near stratified spaces. The first part of the algorithm produces a cover tree using adaptive thresholds based on a combination of multi-scale local principal component analysis and topological data analysis. The resulting cover tree nodes consist of points within or near the same stratum of the stratified space. They are then connected to form a \emph{scaffolding} graph, which is then simplified and collapsed down into a \emph{spine} graph. From this latter graph the stratified structure becomes apparent. We demonstrate our technique on several synthetic point cloud examples and we use it to understand song structure in musical audio data.

\end{abstract}

\section{Introduction}
\label{sec:intro}

We consider \emph{point cloud data}, modeled as a set of points
$\X = \{x_1, \ldots, x_n\}$ in a Euclidean space $\R^D$. Such clouds are hard to analyze directly when $n$ and/or $D$ is large. 
Subsampling techniques are often used in the former situation, and dimension reduction in the latter. In both cases, much care has to be taken to ensure
that the reduction in the number of points and number of dimensions does not destroy essential features of the data. While theorems exist in both contexts, they tend to make assumptions about parameters (like intrinsic dimension) that may be unknown or that may vary widely across $\X$. The latter problem often occurs when $\X$ is not sampled from a manifold, but rather from a \emph{stratified space}.

A stratified space is a topological space that can be decomposed into manifold pieces (called \emph{strata}), of possibly different dimension, all of which fit together in some uniform fashion. A key distinction is between \emph{maximal} and \emph{non-maximal} (also called \emph{singular}) strata: 
briefly, a non-maximal stratum occurs where two or more maximal strata meet. See Figure~\ref{fig:stratifiedExample} for an example. 

This paper proposes a novel, fast, and flexible technique for organizing, visualizing, analyzing, and understanding a point cloud that has been sampled from or near
a stratified space. The technique uses a data structure called the \emph{cover tree} \cite{beygelzimer2006cover} and employs techniques derived from \emph{multi-scale local principal component analysis} (MLPCA, \cite{Le03}) and \emph{topological data analysis} (TDA, \cite{Edelsbrunner2010}). Our method summarizes the strata that make up the underlying stratified space, exhibits how the different pieces fit together, and reflects the local geometric and topological properties of the space.

Instead of subsampling or reducing dimensions, we derive from $\X$ a multi-scale set of graphs, called the \emph{scaffoldings} and the \emph{spine} of $\X$. At each fixed scale, a point in $\X$ belongs to a unique node in these graphs.
A subset of points belongs to a common node only if the local geometry at each of those points is similar enough; that is, only if they belong to a common stratum. We also determine whether a node corresponds to a maximal or non-maximal stratum. In the former case, this suggests that a single dimension reduction technique might be employed on the points in that node.
The edges of these graphs give information about the possible transition regions between different zones of local similarity.

\subsection{Outline of method and paper}
\label{subsec:outline}

After a brief discussion of relevant graph theory (Section~\ref{subsec:graph}), we explain how to represent a point cloud $\X$ by a cover tree $\Tcal$, which is a rooted tree structure with an associated collection of subsets of $\X$ called ``levels."  The points in each level, called ``centers" or ``nodes," are evenly distributed and representative of the point cloud at that scale.
See Section~\ref{subsec:covertree} for more details.

For the next step, described in detail in Section~\ref{sec:pipeline},
we use MLPCA (Section~\ref{subsec:LPCA}) to select a subset of nodes $\Vcal$ from $\Tcal$.
Each node in $\Vcal$ represents a collection of points for which the local structural information (as measured by the eigenvalues of covariance matrices) remains relatively constant. Other uniformity criteria, such as demanding that the points in a node form a single topologically simple cluster, can also be employed. The nodes in $\Vcal$ form the vertices of the scaffolding graph $\Sigma$, whose edge set $\Ecal$ is computed via the geometry of the ambient space. 

Then we label each scaffolding node $v \in \Vcal$ with an estimated local dimension $\Fcal(v)$, using insights from MLPCA and the theory of stratified spaces; this is described in Section~\ref{sec:info}. We note that any other dimension-estimation technique can be used here instead.
Formally, the \emph{scaffolding} of $\X$ is the triple $\Sigma = (\Vcal, \Ecal, \Fcal)$, with $\Fcal: \Vcal \to \{0,1,2,\ldots\}$.
See Figures~\ref{fig:eigenthresholds}-(d) and \ref{fig:biggraph}-(b) for scaffolding examples. 

At this juncture, the collection of points in each fixed node of $\Sigma$ can be fed into a variety of dimension-reduction, topological, or otherwise analytical techniques.
But one can also (Section~\ref{sec:spine}) use graph-collapsing techniques, again informed by insights from the theory of stratified spaces, to produce a much smaller graph
$S= (V,E,F)$, called the \emph{spine} of $\X$. Here a subset of points belongs to a fixed node in $V$ only if those points belong to the same stratum. See Figure~\ref{fig:biggraph}-(c) for a simple example of a spine.

To illustrate this framework, we first run several experiments (Section~\ref{sec:synthetic}) on low-dimensional synthetic datasets, including two that have already been studied in the literature. We also give a high-dimensional example (Section~\ref{sec:music}) with real data, constructing scaffoldings and spines for point clouds created by sliding window embeddings on musical audio.

In all of our synthetic examples, the actual structure of the original stratified space is apparent from the spine graph. We do not yet have a theorem that guarantees that this will always happen, but we close the paper (Section~\ref{sec:discussion}) with a precise conjecture, as well as proposals for future applications to real data.

\subsection{Related work}

Our methods do not compete with dimension-reduction techniques such as principal component analysis \cite{Pearson1901PCA} or manifold learning. Instead, these techniques can be employed at will on the output nodes of our scaffoldings and spines. 

MLPCA has been used before (\cite{BIMNS}, \cite{BIMNS2}, \cite{MLSA}) to analyze data, but not, to our knowledge, in combination with the cover tree.

Many papers (\cite{haro2008translated}, for example) have analyzed stratified spaces as mixtures of general manifolds,
and one \cite{Sayan2014grassman} as a union of flats. Using techniques derived mostly from Gaussian Mixture Models in the former, and Grassmanian methods in the latter, they prove theoretical guarantees about when a point belongs to a maximal stratum, but they do not say much about the singularities. Other work \cite{bendich2012stratlearn} uses topological data analysis to make theoretically-backed inferences about the singularities themselves; however, the algorithms are too slow to be practical.

The cover tree was first introduced in \cite{beygelzimer2006cover} as a way of performing approximate nearest neighbors in $O(\log(N))$ time for a point cloud with $N$ points in arbitrary metric spaces of a fixed dimension (though constants depend exponentially on the dimension).  More recent work has shown that cover trees can be used for analysis of high dimensional point clouds with a low dimensional intrinsic structure \cite{chen2013multi}.  Similarly to \cite{chen2013multi}, we use the cover tree to decompose the point cloud into geometrically simpler parts, but we encode more explicitly the stratified space structure in our representation, and we also autotune which nodes at which levels represent each part, as explained in Section~\ref{sec:pipeline}.

\section{Background}
\label{sec:bground}

%
%
%
%

\subsection{Graph Theory}
\label{subsec:graph}

We give the basic vocabulary about graphs and rooted trees that we will need, the former for defining the spine graph and the latter for the cover tree. For the expert, we note that our graphs are undirected and simple.

A \emph{graph} $G = (V,E)$ consists of a set of \emph{vertices} (or nodes) $V$ and a collection of \emph{edges} $E$, where each edge $e \in E$ is an
unordered pair $e = \{x,y\}$ of vertices. Given such an edge, we say that $x$ and $y$ are \emph{adjacent}, that $e$ \emph{connects} $x$ and $y$, and that $x$ and $y$ are both \emph{incident} to $e$.
The \emph{link} $L(x)$ of a vertex $x$ consists of all vertices adjacent to $x$, and the link of $e = \{x,y\}$ is $L(e) = L(x) \cap L(y)$.
Suppose $f$ is a real-valued function defined on $V$. The \emph{upper link} of a vertex $x \in V$ is $L_{+}(x) = \{y \in L(v) \mid f(y) > f(x) \}$, and the upper link
of the edge $e$ is $L_{+}(e) = L_{+}(x) \cap L_{+}(y).$

Whenever $W$ is a subset of $V$, we can form a new graph $G' = (V', E')$ by \emph{deleting} $W$. Here $V' = V - W$, and $\{x,y\} \in E''$ whenever $x$ and $y$ were both adjacent to some vertex in $W$; all edges from $E$ that connected two non-$W$ vertices are retained.

Alternatively, we can form a different graph $G'' = (V'',E'')$ from $G$ by \emph{collapsing} $W$. In this case, $V''$ is obtained from $V$ by removing $W$ and then adding in a dummy vertex $w$. The edges $E''$ are obtained from $E$ by deleting all edges that connect pairs of vertices from $W$, and also drawing a new edge
from $w$ to each $x \in V - W$ such that $x$ was connected to some vertex from $W$ in $G$.


A \emph{path} of length $m$ between $x$ and $v$ in $G$ is a collection of distinct vertices $x = v_0, \ldots, v_m = v$ such that each $v_i$ and $v_{i+1}$ are adjacent.
A \emph{cycle} is a path from a vertex back to itself.
If there exists a path in $G$ between every pair of vertices, then $G$ is \emph{connected}. Otherwise, $G$ can be decomposed into two or more \emph{connected components}.

A \emph{tree} $T$ is a connected graph with no cycles. If an arbitrary vertex $v \in T$ is distinguished as a \emph{root}, then $T$ becomes a \emph{rooted tree}.
Each non-root node $x$ has a unique path to the root. The $m$th \emph{level} of $T$ consists of all vertices for which the length of this path is $m$. We define a partial ordering on the vertices by declaring $y \leq x$ if and only if the path from $y$ to $v$ passes through $x$, in which case we say that $y$ is a \emph{descendant} of $x$ and
that $x$ is an \emph{ancestor} of $y$. If $y$ and $x$ are one step away from each other, then we use \emph{child} and \emph{parent} instead.

\subsection{Cover Tree}
\label{subsec:covertree}

\begin{figure*}
	\centering
	\subfigure[Level 3]{  \includegraphics[scale=0.6]{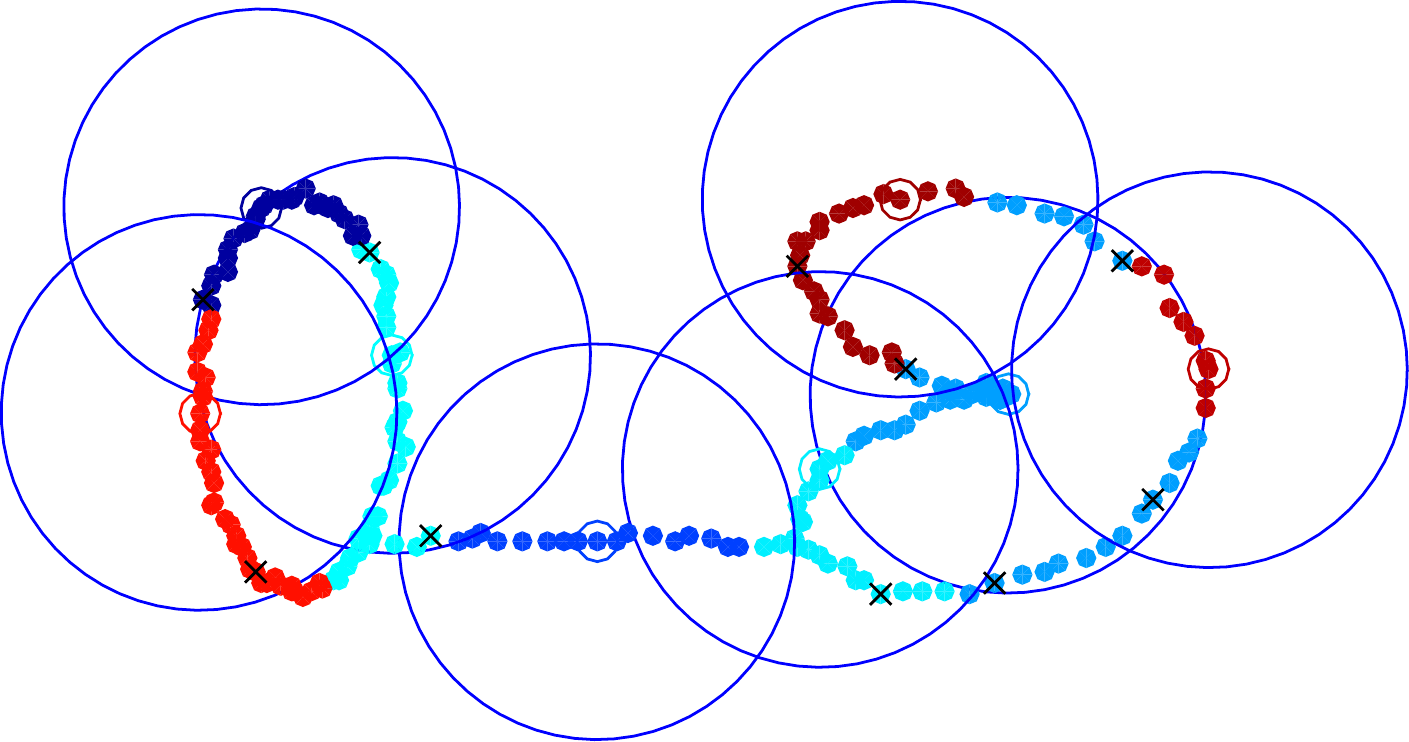} }
	\subfigure[Level 4]{  \includegraphics[scale=0.6]{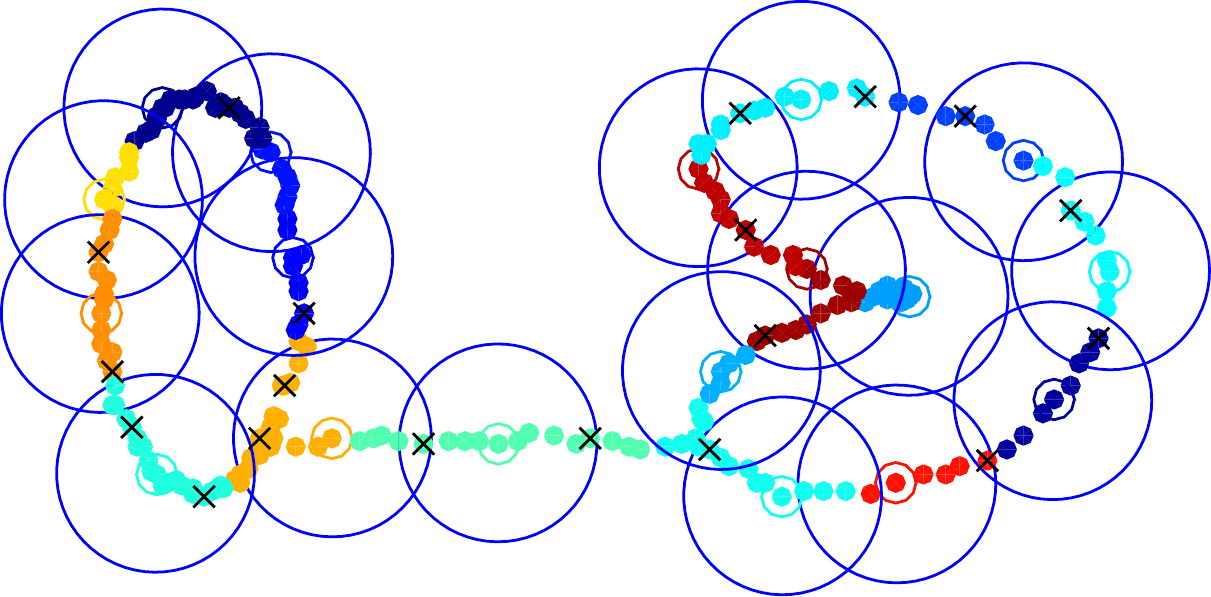} }
	\subfigure[Level 5]{  \includegraphics[scale=0.6]{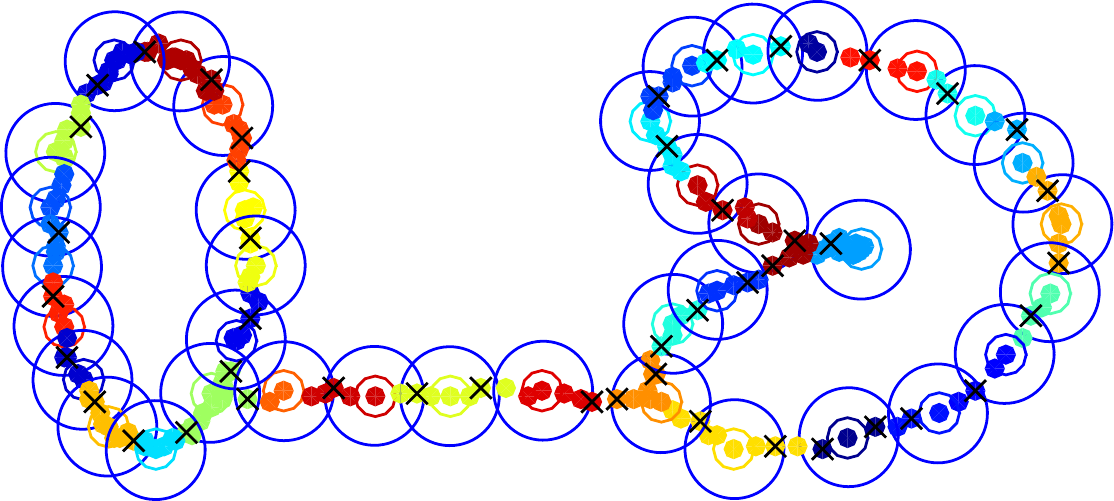} }
	\subfigure[Full Tree Rendering]{  \includegraphics[scale=0.6]{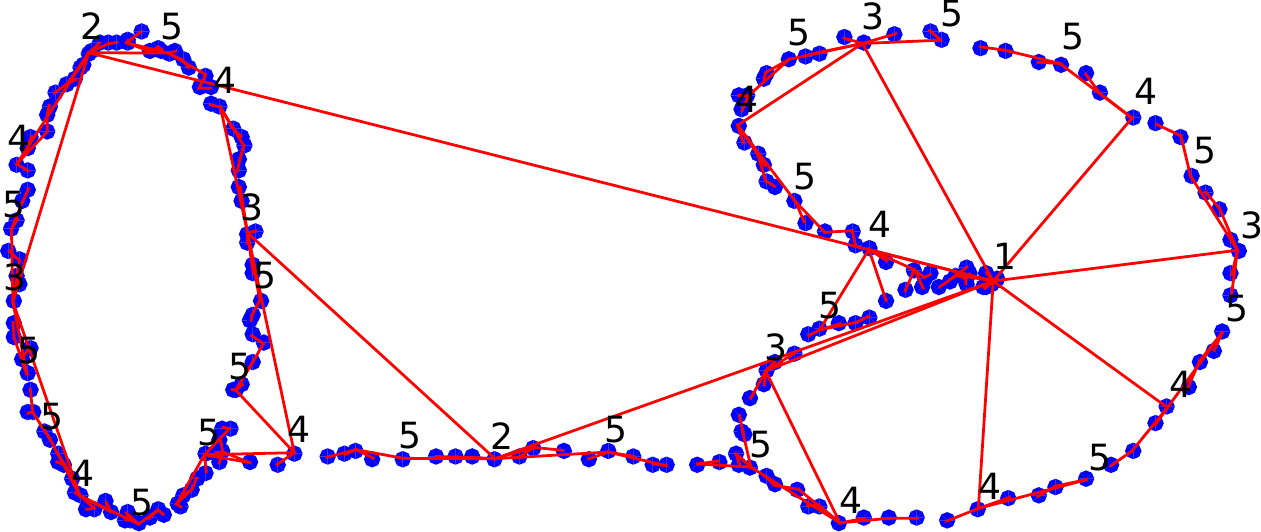} }
	\caption{An example of a cover tree on 250 points in $\mathbb{R}^2$.  In (a)-(c), the centers at level $i$ are circled, and disks with radius $1/2^{i}$ are drawn in blue around each center.  Points within the same subtree rooted at a center at that level are drawn with the same color.  Centers at level $i+1$ are marked with an X to illustrate the covering condition.  The tree is drawn in (d), with points in the first five levels numbered with the level in which they first occur.  }
\label{fig:CoverTreeExample}
\end{figure*}

\begin{definition}
A \textbf{cover tree} $\Tcal$ on a point cloud $\X$ with a metric $d_{\X}$ is a leveled tree on $\X$ where each node is associated to one of the points in $\X$ (called ``centers''), and where the collection of $C_i$ (set of points in $\X$ that are level $i$ centers for $i \in \mathbb{Z}$) satisfy the following conditions:
\begin{itemize}
\item (Nesting) $C_{i} \subset C_{i+1}.$  
\item (Covering): For each $C_{i+1}^j \in C_{i+1}$, there exists a node $y \in C_{i}$ so that $d_X(C_{i+1}^j, y) \leq R_i$, where $R_{i}=1/2^{i}$.  
\item (Packing): 
For all $ C_i^j \neq C_i^k \in C_{i}$, $B_{R_{i}}(C_i^j)\cap B_{R_{i}}(C_i^k) = \emptyset$, where $R_{i}=1/2^{i}.$  
\end{itemize}
\noindent Here, $C_i^j$ indicates the $j$th center at level $i$.
\end{definition}

Another way of phrasing the covering condition is to say that balls of radius $R_i$ centered at the points in $C_i$ will cover the whole point cloud. Exactly one such $y \in C_i$ is designated as the parent of $C_{i+1}^j$ in the tree structure.  The packing condition promotes levels which contain equally spaced centers.  Note that in this definition, it is possible for a node at one level to be its own parent one level up. 
There is a minimum $i$ for which the size of $C_i$ is $1$, and the node at that value of $i$ is designated as the root node.  Similarly, there is a maximum $i$ for which $C_i = \X$.  Figure~\ref{fig:CoverTreeExample} shows an example cover tree on a point cloud in $\mathbb{R}^2$.

\subsection{Stratified Spaces}
\label{subsec:SS}

\begin{figure}[ht]
\includegraphics[scale=0.11]{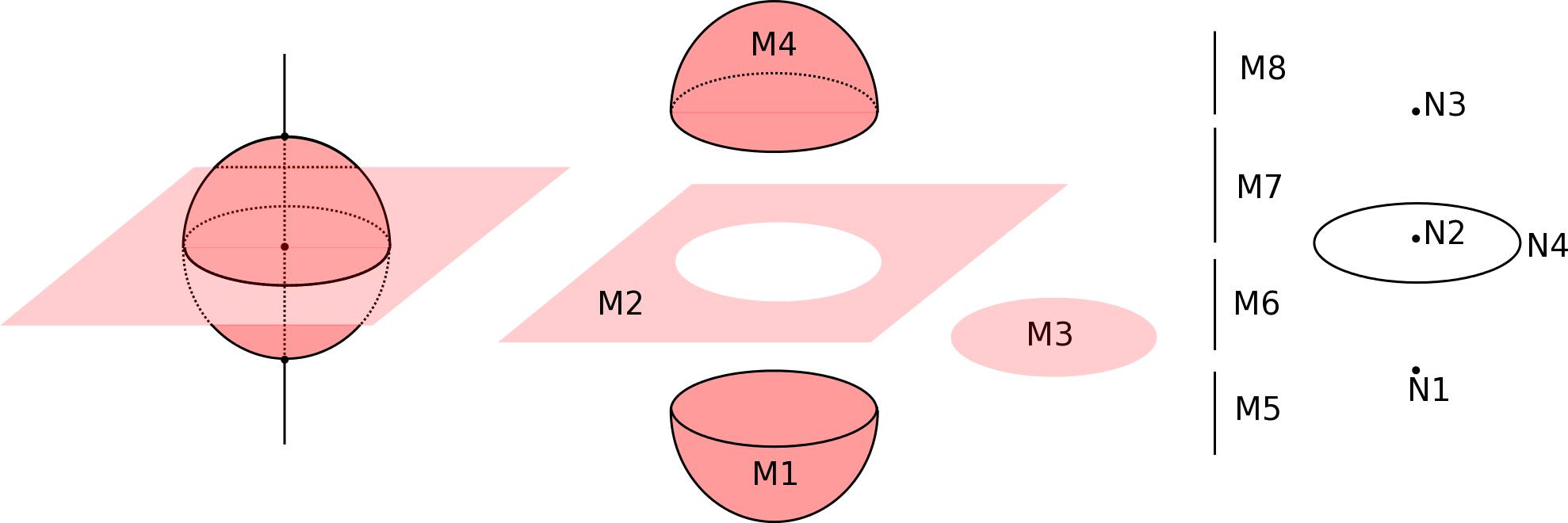}
\caption{Far left: A simple stratified space $Y$. Rest of figure: a stratification of $Y$ into $12$ strata; $M_1$ through $M_8$ are maximal,
and $N_1$ through $N_4$ are non-maximal. }
\label{fig:stratifiedExample}
\end{figure}

Recall that a space $Y$ is called a \emph{manifold} of dimension $m$ if every point in $Y$ has a neighborhood that is homeomorphic to $\R^m$. 
The space $Y$ on the left of Figure~\ref{fig:stratifiedExample}, consisting of a sphere intersecting with a horizontal plane at its equator and with a line piercing
it at the north and south poles, is not a manifold: for example, every small enough neighborhood of the north pole looks like a plane pierced by a line. 
A \emph{stratification} of $Y$ is a decomposition of $Y$ into connected manifolds $\{S_i\}$ called \emph{strata}. Each stratum is required to fit nicely
within the whole of $Y$; see \cite{Hughes2000surgery} for rigorous versions of this condition and this entire discussion.
The rest of the figure shows a stratification of $Y$.

A stratum $S_i$ is called \emph{maximal} 
if it is disjoint from the closure of any other stratum. The other strata are called \emph{non-maximal}.
If $z$ is a point on a non-maximal stratum, then every small neighborhood of it within $Y$ looks like the intersection of two or more strata.
In our example, the maximal strata are labeled $M_1$ through $M_8$; note that there is no reason for the maximal strata to all have the same dimension. Each of the non-maximal strata, $N_1$ through $N_4$, belongs to the closure of two or more of the maximal strata.


\subsection{Local Principal Component Analysis}
\label{subsec:LPCA}

We assume the reader is already familiar with principal component analysis (PCA, \cite{Pearson1901PCA}). 
Here we regard PCA simply as a machine that takes in a point cloud, computes a covariance matrix, and returns the eigenvectors 
along with the corresponding eigenvalues given in non-increasing order. This ``eigeninformation'' may then be used for dimensionality reduction. 

\begin{figure}
\centerline{\fbox{\includegraphics[scale=0.21]{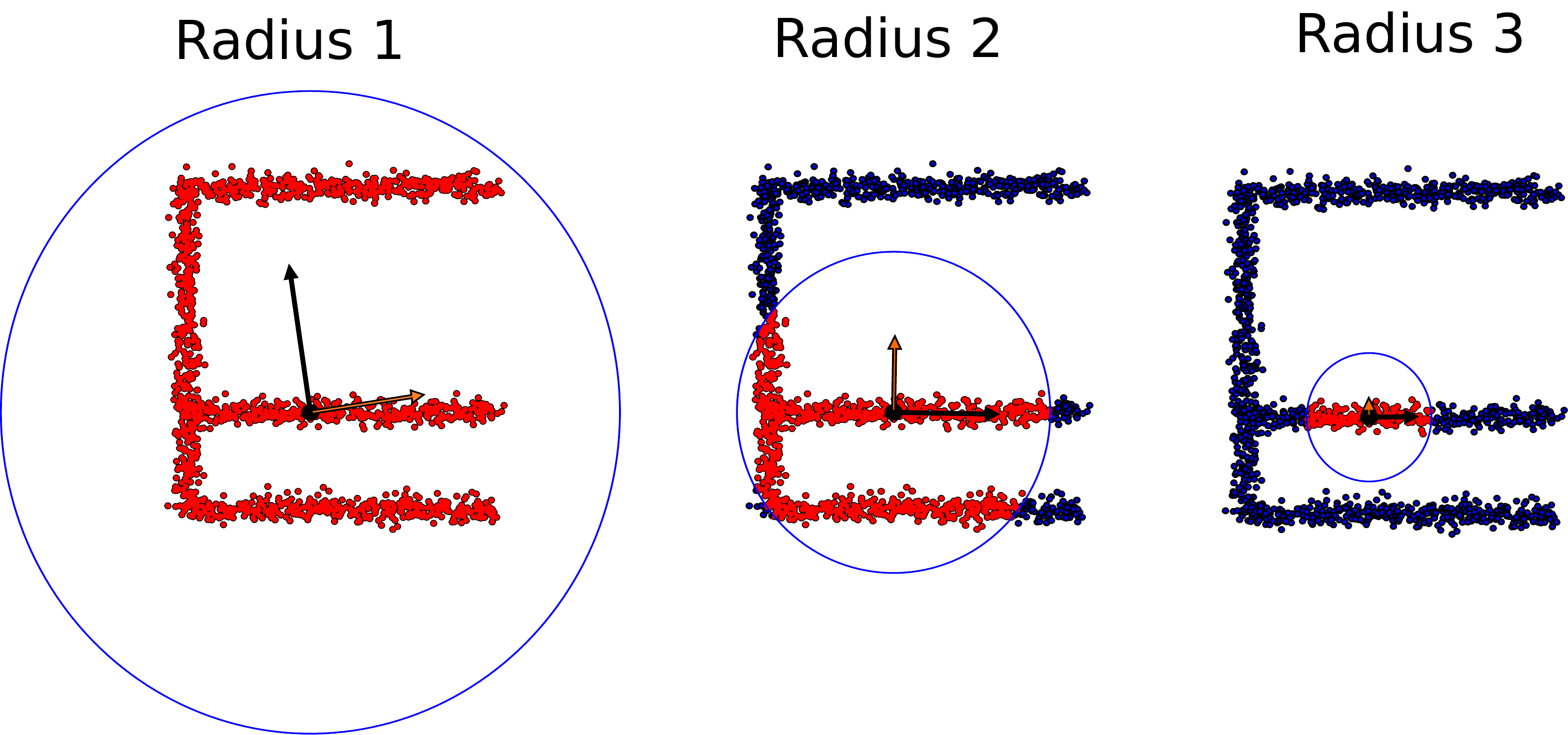}}}
\caption{Illustration of the technique of multi-scale local principal component
analysis (MLPCA).  Directions of largest variance are drawn as black arrows, and directions of smallest variance are drawn as orange arrows in this 2D example } \label{fig:MLPCA}
\end{figure}

The technique of \emph{multi-scale local principal component analysis} (MLPCA \cite{Le03}), takes as input a point cloud $\X \subset \R^D$, a particular point $p \in \R^D$
and a radius $R$, and returns as output the results of PCA run only on the points in
$\X$ that lie within the Euclidean $R$-ball around $p$. This is typically done at multiple radius scales and at many center 
points; see Figure~\ref{fig:MLPCA}.

We close with an observation that will be crucial later.
Suppose the point cloud $\X$ has been sampled from a stratified space $Y$. After computing MLPCA at a few radii around a specific point $p \in \X$, one might consistently see $k$ dominant eigenvectors, and thus try to conclude that the ``local dimension'' around $p$ is $k$; in other words, that $Y$ looks locally
like a $k$-flat near $p$. This is certainly possible, but need not be the case. Instead, $p$ might be near a non-maximal stratum of $Y$, and the local picture might be more like the intersection of an $\ell$-flat and a $k - \ell$ flat, for some non-zero value of $\ell$. 

\subsection{Persistent Homology}
\label{subsec:PH}

Algebraic topology associates to a topological space $X$ a series of abelian groups $H_i(X)$,
called the \emph{homology groups} of $X$, which measure its higher-order connectivity.
The ranks of these groups are called 
the \emph{Betti numbers} $\beta_i(X)$:
$\beta_0$ counts the number of connected components of $X$, $\beta_1$ (resp. $\beta_2$) counts
the number of independent non-bounding loops (resp., closed surfaces) in the space, and so forth.
See \cite{Munkres2}, for example, for more precise definitions.

\emph{Persistent homology} transforms these algebraic notions into a 
measurement tool relevant to high-dimensional and noisy data.
The key idea is that we are rarely interested in the Betti
numbers of a particular fixed space; rather, we are concerned with the homological features
that persist across a wide interval in a one-parameter filtration of a space, and we represent
these features in a compact visual way as a \emph{persistence diagram} in the plane.
For precise definitions and statements, see for example \cite{Edelsbrunner2010} or \cite{Chazal2009b}.

\paragraph{Multi-scale clustering}

\begin{figure}
 \begin{center}
  \includegraphics[scale=0.25]{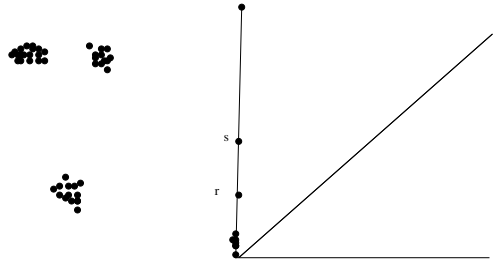}
 \end{center}
\caption{A point cloud $\X$ (left) and the zero-dimensional persistence diagram (right). The values $r$ and $s$ are half
the distance between the two top clusters and between the right and bottom cluster, respectively.}
\label{fig:cluster}
\end{figure}

The simplest homological invariant of a space $X$ is $\beta_0(X)$, its number of components. Let $\X$ be the point cloud seen on the left of Figure \ref{fig:cluster}. 
The most literal interpretation of $\beta_0(\X)$ here would be $N$, where $N = |\X|$.
On the other hand, most off-the-shelf clustering algorithms would report that $\X$ has three components, since
that seems to be the most natural way to group the points.
Some algorithms might report back two clusters (grouping the top two into one), depending on how parameters are chosen.
Persistent homology in dimension zero gives a useful description of all of these answers
along with a report of the scales at which they are valid.

More precisely, let us define, over a set of \emph{scales} $\alpha \geq 0$, the space $\X_{\alpha}$ to be the union of closed balls of radius $\alpha$ centered
at each point in the cloud.
As $\alpha$ increases, we imagine the point cloud gradually thickening into the ambient space, and more and more components begin to merge into
one another.
An $\alpha$-value at which a merger occurs will be called a \emph{death}, and we record the multi-set of these deaths as dots in the zero-dimensional
\emph{persistence diagram} of $\X$, as shown on the right side of Figure \ref{fig:cluster}.
One way to translate between this diagram and a traditional number-of-clusters answer is to cut the diagram by a horizontal line, and then report the number
of dots above that line. This shows that the answers ``two clusters'' and ``three clusters'' are both quite reasonable, since each could be obtained
via a wide range of threshold parameters.
On the other hand, the answer ``nine clusters'' is in no way stable, since there is only a very narrow range of threshold parameters that would report that number.

\paragraph{Higher homology of point clouds.}

In general, the $k$-dimensional persistence diagram $D_k(\X)$ of a point cloud will summarize the evolution
of the $k$-dimensional homology group $H_k(\X_{\alpha})$, while $\X$ thickens as described above.

\begin{figure}
 \begin{center}
  \includegraphics[scale=0.13]{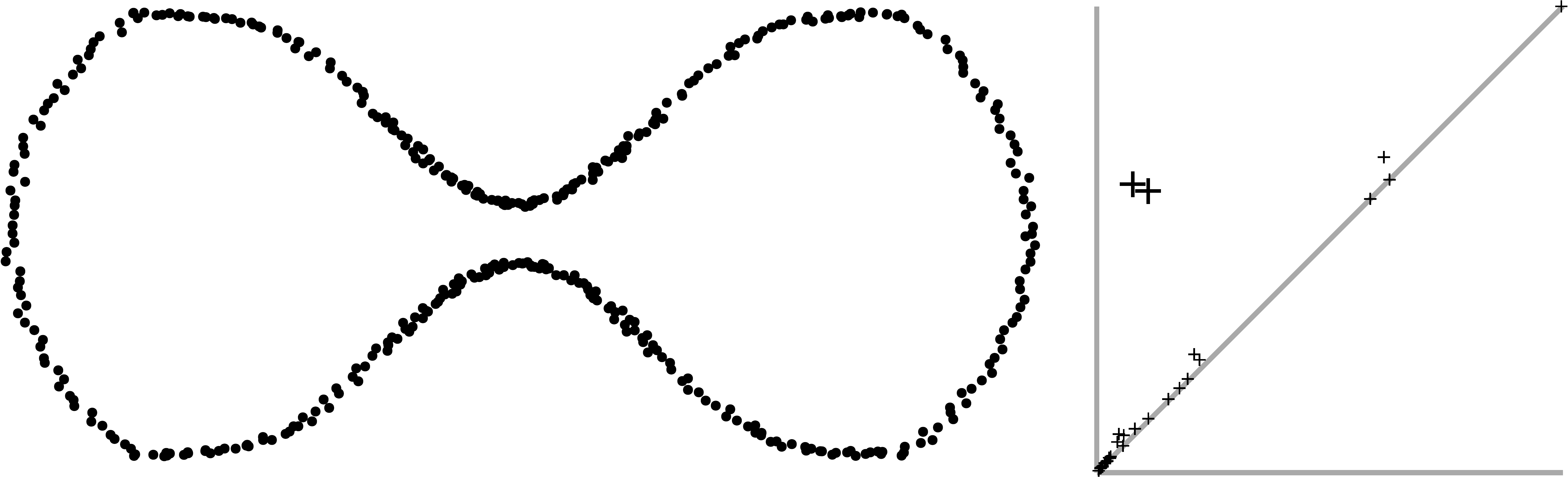}
 \end{center}
\caption{A point cloud sampled from a closed curve in the plane (left) and the persistence diagram in dimension one (right).}
\label{fig:eight-sketch3}
\end{figure}

Consider the left side of Figure \ref{fig:eight-sketch3}, which shows a point cloud $\X$ sampled, with some noise,
from a closed curve in the plane.
As $\X$ starts to thicken, all of the components very quickly collapse into one. 
However, there is also a new phenomenon that occurs: namely, the formation of a closed loop.
We call this the \emph{birth} of a one-cycle.
When $\alpha$ increases enough to build a bridge across the neck of the shape, a new one-cycle is born.
Each of these one-cycles is said to \emph{die} at the $\alpha$-value for which it is possible
 to contract
the loop to a point within $\X_{\alpha}$.
The (birth, death) pairs corresponding to the lifetime of each cycle are stored in $D_1(\X)$, shown
on the right side of the same figure. 
The \emph{persistence} of a dot $u = (x,y)$ is $y - x$.

Suppose that $\X$ is a good sampling from a compact space $Y$.
There are then a series of ``homology inference'' theorems (\cite{CohenSteiner2007}, \cite{Chazal2009b}) which prove that one can
read off the ``ground truth'' homology of $Y$ by examining the points in $D_k(\X)$ which sit, very roughly speaking, near the $y$-axis
and are of high-enough persistence.

\section{The Scaffolding}
\label{sec:pipeline}


Given a point cloud $\X \subseteq \R^D$, one can let $d_{\X}$ be the Euclidean metric and build the full cover tree $\Tcal$ as above. Instead, at each node in the cover tree, we use a criterion to decide whether we will continue to construct the subtree rooted at that node, or whether all of the points contained in that subtree are sufficiently ``uniform.'' In the latter case, the entire subtree is compressed to a single node. At the end of this process, we output the leaves as $\Vcal$, the nodes of our scaffolding graph.

\subsection{Eigenmetric threshold}
\label{subsec:CTconstruct}

Our adaptive tree building construction is based on MLPCA, as we now explain.
First, we choose some increasing set of radii $r_1, \ldots r_n$. For each point $p \in \X$ and each $i =1, \ldots n$,
let $\boldsymbol \lambda^i (p) = (\lambda_1^i, \lambda_2^i, \ldots \lambda_D^i)$ be the eigenvalues, in non-increasing order, that result from performing MLPCA on $\X$ with radius $r_i$ and center point $p$.
Then for any pair of points $p,q \in \X$, we define
\begin{equation}
\label{eq:eigenmetric}
E(p,q)=||(||\boldsymbol \lambda^1 (p)-\boldsymbol \lambda^1 (q)||,\ldots,||\boldsymbol \lambda^n (p)-\boldsymbol \lambda^n (q)||)||,
\end{equation}
\noindent where $||\cdot||$ is the usual Euclidean norm. We shall refer to $E$ as the \emph{eigenmetric} on $\X$. Points with similar ``eigenprofiles'' across multiple scales will have small eigenmetric distance. Note that $E$ is a pseudometric as $E(p,q)$ may be 0 for $p \neq q$ (e.g., consider two copies of the same point cloud laid out side by side).

The eigenmetric is used to construct $\Vcal$ as follows. Fix a level $i$ of $\Tcal$. At each node $C_i^j$ on this level, we perform an eigenmetric threshold test on the points from $\X$ associated with $C_i^j$ that would end up in its subtree upon further construction.  To do this, we look at the points associated with each child of $C_i^j$, and we only further subdivide a node when the eigenmetric between a pair of children exceeds an \emph{eigenthreshold} $\tau$. Otherwise, we stop constructing the subtree below $C_i^j$ and instead add $C_i^j$ to $\Vcal$.
At the end of this process, every point in $\X$ belongs to a unique node in $\Vcal$. We refer to the set of points belonging to a given node $v$ as the \emph{cluster} of $v$.

\begin{figure*}[ht]
(a)\includegraphics[scale=0.14]{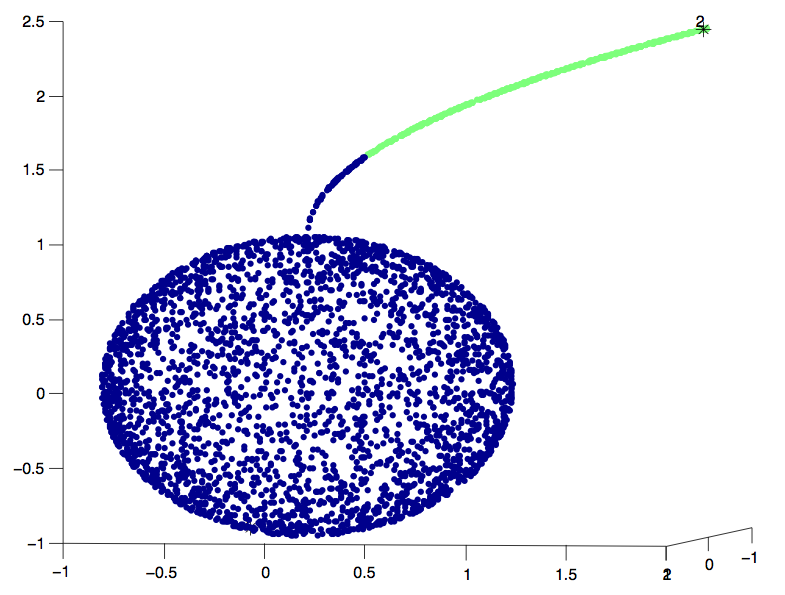}(b)\includegraphics[scale=0.15]{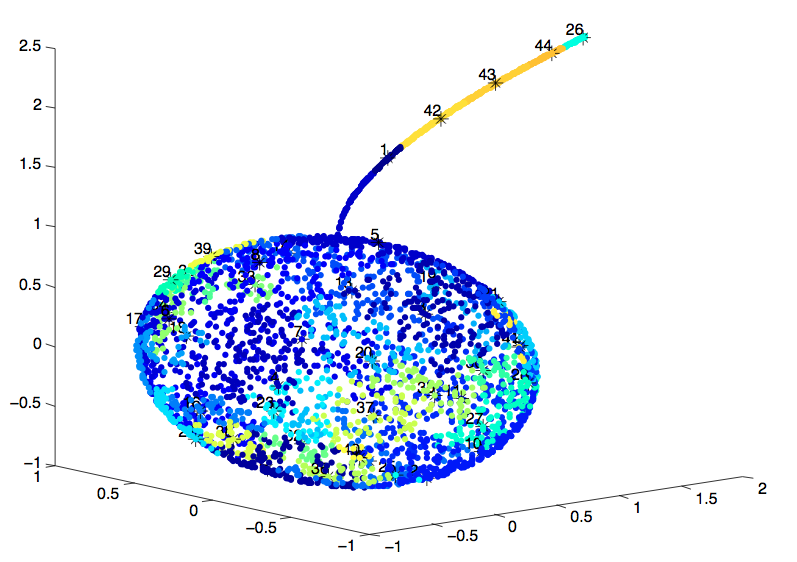}(c)\includegraphics[scale=0.15]{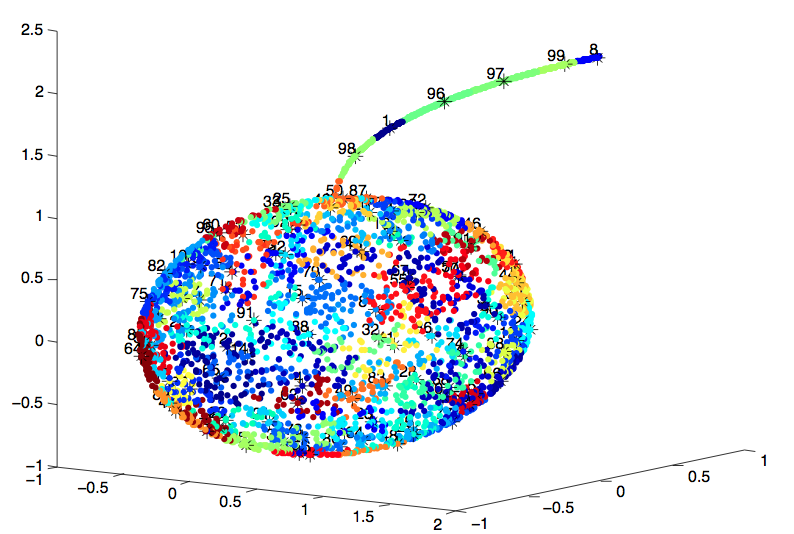}
(d)\includegraphics[scale=0.15]{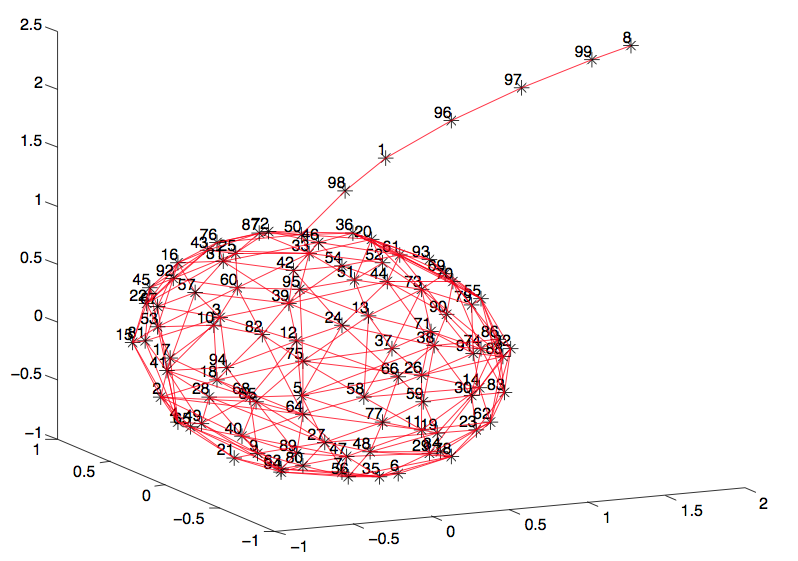}
\caption{Results of the adaptive cover tree construction corresponding to eigenthresholds (a) $\tau = 1$, (b) $\tau = 0.1$, and (c) $\tau=0.001$ for a point cloud sampled from a sphere with an adjoining curve. Points belonging a single $\Vcal$-node are given the same color. Also shown is (d): A scaffolding for the last cover tree with automatically determined distance threshold $\delta \approx 0.1$.}
\label{fig:eigenthresholds}
\end{figure*}

Parts (a)-(c) of Figure~\ref{fig:eigenthresholds} show this process. Note that one must choose $\tau$. To reap the multi-scale benefits, one should construct the scaffolding nodes for a range of eigenthreshold values and witness how they change.

\subsubsection*{Additional subdivision criteria} 

It may be desirable to have the set of points belonging to a specific node in $\Vcal$ satisfy a variety of ``simplicity'' properties: for example,
one might want them to look like one cluster rather than many. 
This can be achieved by introducing a second criterion based on a choice of multi-scale clustering technique. For example, one might compute the zero-dimensional persistence diagram on the set of points in $\Vcal$ to quantify how much thickening is needed before these points become one component. If this value exceeds a user-specified threshold, the node is further subdivided. 
We follow this practice in some of our examples below. 

One could go even further, and demand that the points in a node look genuinely like a contractible clump, rather than having any large voids or holes. This can be done by computing the higher-dimensional persistence diagrams on the points belonging to the node, and then subdividing that node if any dots of persistence higher than a user-specified threshold appear in any of the diagrams. To make this computationally feasible, one can leverage recent approximation schemes \cite{sheehy13linear} for persistent homology that involve the cover tree.

\subsection{Building the scaffolding}
\label{subsec:scaffolding}

Now we describe how to build the scaffolding graph $\Sigma$ on the nodes $\Vcal$.
A pair of nodes $v,w \in \Vcal$ forms an edge in $\Ecal$ if and only if the Euclidean distance between $v$ and $w$ is below some threshold. The value $\delta$
of this threshold can be chosen manually, or one may use a variety of automatic procedures based on prior assumptions.
For example, if one feels that $\X$ comes from an underlying connected space, then one might desire that $\Sigma$ be connected.
To enforce this, one can use zero-dimensional persistent homology on $\X$ and let $\delta$ be the smallest thickening value that creates one component
from $\X$. To avoid choosing a distance threshold, one may select a range of values for $\delta$ and watch how the scaffolding changes as the distance threshold varies.
See part (d) of Figure~\ref{fig:eigenthresholds} and part (b) of Figure~\ref{fig:biggraph} for examples of scaffolding graphs. 



\section{Local Dimension and other Data}
\label{sec:info}

\subsection{Local Dimension Estimation}
\label{subsec:lpca}

Once the scaffolding $\Sigma$ is constructed, we compute a nonnegative-integer-valued function $\Fcal$ on $\Vcal$.
If the cluster of a node $v$ comes from well within a maximal stratum, then $\Fcal(v)$ estimates the local dimension of that stratum.
We describe our choice of $\Fcal$ here, but point out that any number of other intrinsic dimension estimation techniques (for example, the box-counting method of \cite{haro2008translated}) can be substituted into our essential framework. 

For each node $v$, we begin by performing PCA on a point set made up of the union of the cluster of $v$ and all clusters corresponding
to neighbors of $v$ in $\Sigma$. We compute the square roots of the eigenvalues from PCA, normalizing so that their values range between 0 and 1.  We then compute the differences between successive eigenvalues, including the difference between the smallest eigenvalue and 0, and use the location of the largest ``eigengap'' as our initial estimate of local dimension near that cluster. 

\subsubsection*{Refining dimension estimates}

Next, we refine these initial estimates by building in knowledge of maximal vs. non-maximal strata in generic stratified spaces. 
Specifically, we consider the nodes $w \in \Vcal$ which have at least one neighbor $x \in L(w)$ such that $\Fcal(x) < \Fcal(w)$.
If $w$ belongs to the set $\Wcal$ of such nodes, then it is possible that the cluster of $w$ was sampled from a non-maximal stratum.
This is due to the fact that, when two or more maximal strata come together, PCA will see the dimension near the intersection as the sum of the dimensions
of the individual maximal strata.

Now if $w \in \Wcal$ truly represents a non-maximal stratum where two or more maximal strata come together to create a singularity, then (1) its link $L(w)$ in the scaffolding should be disconnected (with the connected components corresponding to the nearby maximal strata) and (2) the link should include nodes whose dimensions ($\Fcal$-values) sum to at least its estimated dimension, i.e., 
\begin{equation}
 \Fcal(w) \leq \sum_{x \in L(w)} \Fcal(x).
\end{equation}
If either of these conditions fails to hold, we conclude that the $\Fcal$-values in $L(w)$ were miscalculated and relabel them all with $\Fcal(w)$.

\begin{figure}
(a) \includegraphics[scale=0.13]{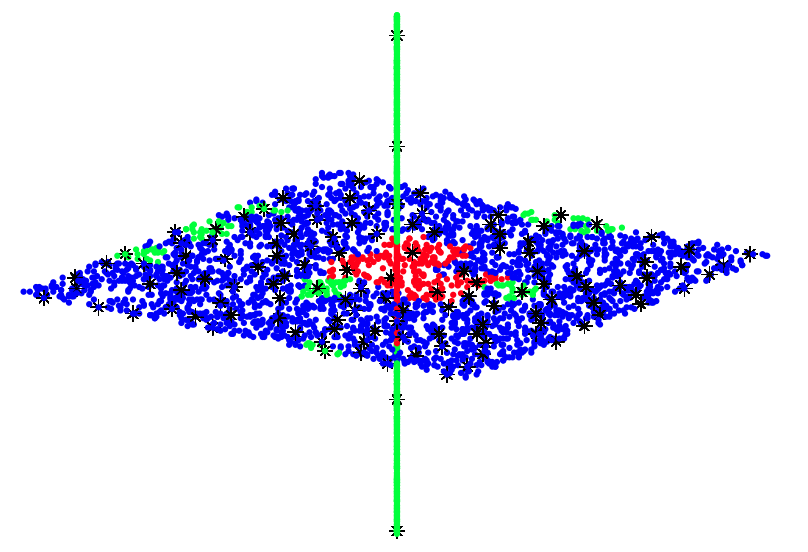} (b) \includegraphics[scale=0.13]{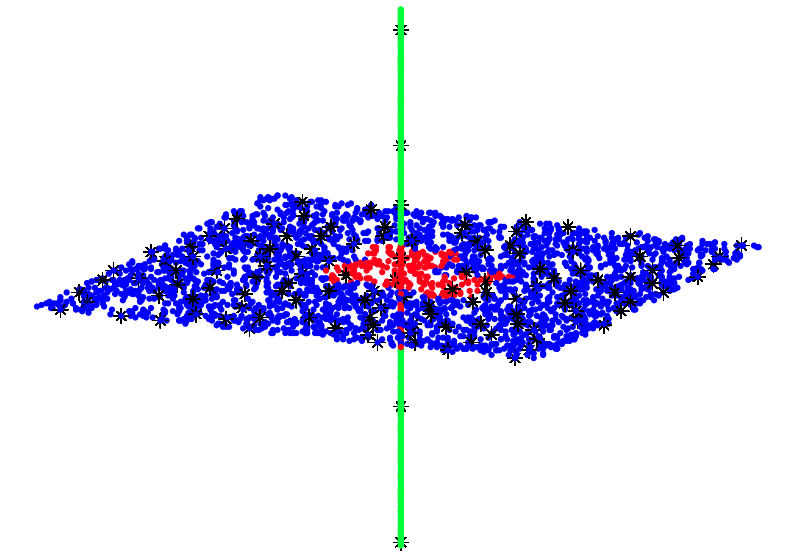}
\caption{Results of local dimension estimation prior to refining the estimates (a) and after the refinement process (b).}
\label{fig:refining}
\end{figure}

Figure~\ref{fig:refining} shows an example, with points sampled from a plane intersecting a line in $\R^3$.
The nodes $\Vcal$ of the scaffolding $\Sigma$ are shown as black asterisks, and the local dimension estimation is illustrated by color in the figure. Red points belong to nodes labeled 3-dimensional, and since such nodes have neighbors in the scaffolding with smaller dimensions, they are candidates for non-maximal strata. Green points scattered on the plane belong to nodes labeled 1-dimensional, so the blue 2-dimensional nodes that are connected to such 1-dimensional nodes are also candidates for being non-maximal strata. After refining the estimates using the theory above, the 1-dimensional nodes scattered on the plane are determined to be errors in the refining process, and are therefore relabeled with improved dimension estimates.
Parts (a) and (b) show the estimates before and after refinement, respectively.

The local dimension estimate process is also a multi-scale process. It is possible that performing PCA on a node and its neighbors in the scaffolding at a small distance threshold does not give an accurate representation of the local geometric structure near the node. Perhaps there are too few points to capture the local picture, or perhaps the shape of the clusters is such that the dimension estimation from PCA would result in an error. To improve one's understanding of the intrinsic local dimension, it is best to employ a generalized version of MLPCA in the local dimension estimation process. That is, one may preserve the value of $\delta$ for the scaffolding but gradually relax the threshold for including more clusters in the PCA computations for each node.  For example, for the point cloud in the top of Figure~\ref{fig:biggraph}, although the distance threshold choice of $\delta=0.05$ yields a connected scaffolding graph, the dimension estimation process of performing PCA on a cluster and those clusters connected to it in the scaffolding yields a too-local understanding of the dataset for this choice of $\delta$. The results of the local dimension estimation that are shown in Figure~\ref{fig:biggraph} were the result of computing, for each cluster, PCA on the clusters within a distance threshold range of approximately 0.12 to 0.23.

\subsubsection*{Topology of strata} As an option to understand the topology of the strata, one can perform persistence on the points --- or better, on only the node centers --- corresponding to each dimension-based connected component in the scaffolding. One can then decorate the nodes with vectors of persistent Betti numbers $(\beta_0,\beta_1,\beta_2,\ldots)$; see Figure~\ref{fig:popspine} in Section~\ref{sec:synthetic}. We point out that due to the flexible nature of our basic framework, one could also apply a different manifold learning technique of one's choice to obtain information regarding the topology of the strata.


\section{The Spine}
\label{sec:spine}

Local dimension estimation works best when there are many cover tree nodes, which is a result of choosing a small eigenthreshold. Consequently, the scaffolding
$\Sigma$ has many vertices. We now describe a further procedure that radically simplifies $\Sigma$ to produce a new graph $S$ on a much smaller vertex set.
This graph, called the \emph{spine} of $\X$, provides a streamlined and efficient summary of the strata that make up the stratified space which produced $\X$.

\begin{figure*}[ht]
(a) \includegraphics[scale=0.42]{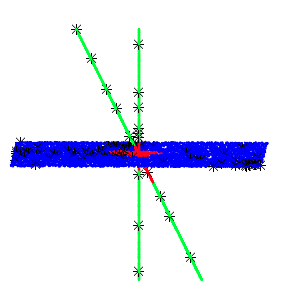}
(b) \includegraphics[scale=0.5]{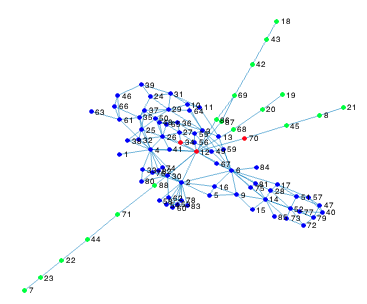}
(c) \includegraphics[scale=0.37]{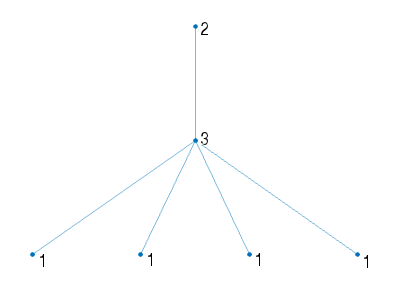}
\caption{Local dimension estimation on a point cloud (a), as well as the scaffolding (b) and spine (c), for eigenthreshold $\tau=0.1$, a persistent $H_0$ threshold of 0.05, and distance threshold $\delta = 0.05$. Asterisks indicate node centers and green, blue, and red points indicate 1-, 2-, and 3-dimensional areas, respectively.}
\label{fig:biggraph}
\end{figure*}
Using the procedure above, $\Vcal$ is partitioned into maximal nodes $\Mcal$ and non-maximal ones $\Ncal$.
We then further partition $\Ncal$ by $\Fcal$-value and locate the connected components therein.
These connected components are then collapsed into single nodes (as described in Section~\ref{subsec:graph}). Note this retains all edges that previously connected these components to nodes of other dimensions.


Now we process the maximal nodes $\Mcal$, starting by partitioning them by $\Fcal$-value and then taking connected components within this partition.
Let $\Ccal$ be one such component. We divide $\Ccal$ further into \emph{boundary} nodes and \emph{interior} nodes, where a node is
in the boundary if it is connected to at least one node of a different dimension.
The interior nodes are deleted (as described in Section~\ref{subsec:graph}). 
Then, for any edge $e = \{x,y\}$ that connects two boundary nodes, we check whether $L_{+}(x) = L_{+}(y)$; in other words, we check whether $x$ and $y$ are connected
to the same set of non-maximal nodes. If so, then the edge $e$ is collapsed.
Running through this process for each component of $\Mcal$ produces
the vertices $V$ and edges $E$ of the spine $S$. We note that vertices with different $\Fcal$-value never get collapsed together. Therefore, it makes sense to label each node in $V$ with the $\Fcal$-value of any node that was collapsed into it. This labeling produces $F$, and we have our spine
$S = (V,E,F)$.

For example, the scaffolding $\Sigma$ for the point cloud in Figure~\ref{fig:biggraph} (a) is in part (b) of the same figure.
If we run this process on $\Sigma$, we obtain the spine $S$ in part (c). Note that $S$ is in fact the Haase diagram for the partial ordering on the strata of a plane
pierced by two lines (see Section~\ref{sec:discussion}).




\section{Synthetic Examples}
\label{sec:synthetic}

This section demonstrates our algorithms on several synthetic examples\footnote{code to run these examples appears at https://gitlab.com/elleng/GeometricModelsCode}.

\begin{figure*}[h]
(a) \includegraphics[scale=0.275]{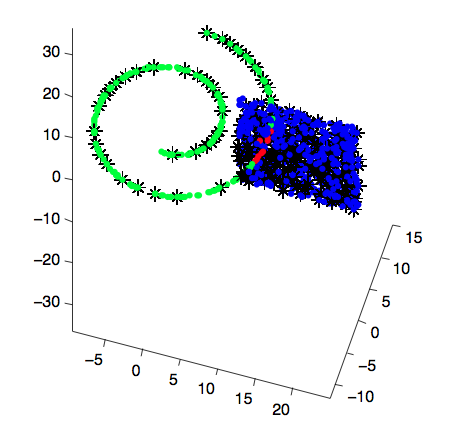} (b) \includegraphics[scale=0.375]{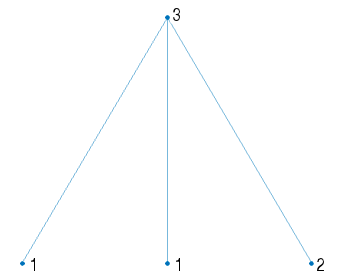}
(c) \includegraphics[scale=0.2]{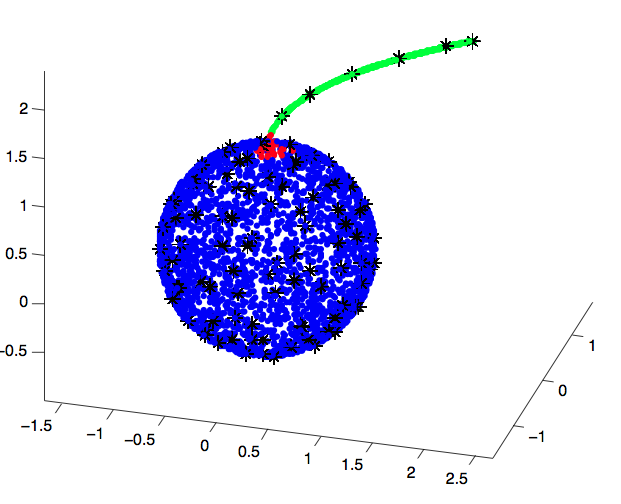} (d) \includegraphics[scale=0.45]{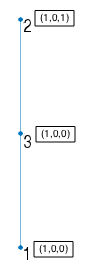}
\caption{(a) Local dimension estimation and (b) spine for the spiral-plane example. (c) Local dimension estimation and (d) spine for sphere-curve example.}
\label{fig:spiralPlane}
\end{figure*}

We begin by considering two synthetic examples that appeared in previous related work. First, as in \cite{haro2008translated} and \cite{chen2013multi}, we sampled 300 points from a spiral and 800 points from a plane in $\R^3$; the spiral and plane intersect transversely. We constructed the adaptive cover tree and scaffolding for threshold choices $\tau=0.1$ and $\delta=2.5$. Part (a) of Figure~\ref{fig:spiralPlane} shows the refined local dimension estimation, and part (b) displays
the spine. The red points, labelled as 3-dimensional, indicate the presence of a non-maximal stratum in the underlying stratified space: namely, the singular stratum where the two 1-dimensional spiral pieces intersect the 2-dimensional plane. The spine captures how the strata are situated relative to one another.

Our second example, a sphere with an attached curve-fragment, also appeared in \cite{haro2008translated}. Part (c) of Figure~\ref{fig:spiralPlane} shows the refined local dimension estimation and part (d) shows the spine. Again our method identifies the important location where two maximal strata of different dimensions meet. We point out that the same spine would have been produced by a point cloud sampled from a $2$-plane with a ray sticking out of it. Persistent homology, computed on each node, distinguishes these two cases. In part (d), the spine nodes are decorated with vectors $(\beta_0,\beta_1,\beta_2)$ of persistent Betti numbers.

\begin{figure}[ht]
\begin{center}
(a) \includegraphics[scale=0.35]{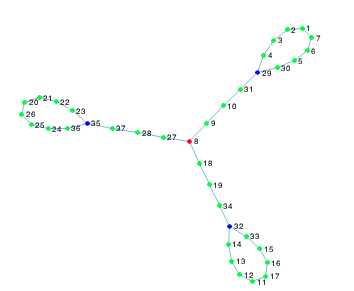} \\
(b)\includegraphics[scale=0.35]{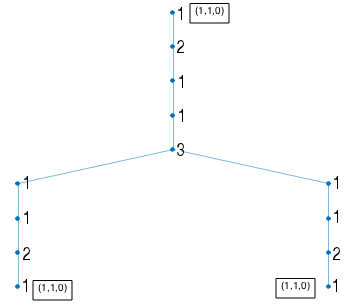}
\caption{(a) The scaffolding and (b) spine for a point cloud sampled from three lollipop-like structures in $\mathbb{R}^6$ ($\tau=0.001$ and $\delta = 0.1$).}
\label{fig:popspine}
\end{center}
\end{figure}

For another example that incorporates topological information, we constructed a point cloud sampled from three lollipop-like structures in $\mathbb{R}^6$ (6000 points in total) belonging to three distinct, mutually orthogonal planes in $\R^6$. Each lollipop-like structure is made up of a circle and a line segment emanating out from a point on the circle, and the three line segments meet at a single point at each of their bases. See Figure~\ref{fig:popspine}. Note that the three-dimensional non-maximal node corresponds to the point of contact of the three line segments, and the two-dimensional non-maximal nodes occur at the intersections of the circles and line segments. Every node consists of a single persistent connected component. Some nodes have been decorated with vectors of persistent Betti numbers, and these are the nodes with non-trivial 1-dimensional homology. A non-decorated node can be assumed to have persistent homology described by the vector $(\beta_0,\beta_1,\beta_2)=(1,0,0)$.

Note that our rendering preserves an essential topological feature: the actual loops. A linear technique, such as PCA, will often completely destroy features like this.

\begin{figure}[ht]
\centering
\includegraphics[scale=0.4]{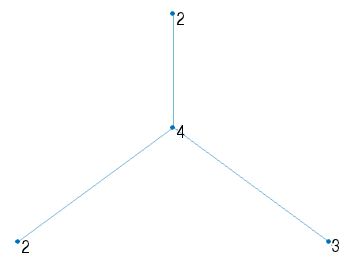}
\caption{The spine for a point cloud sampled from a $2$-plane and $3$-plane intersecting along a line in $\R^4$.}
\label{fig:2planesspine}
\end{figure}

Finally, for our last synthetic example, we consider a 2-plane and a 3-plane intersecting along a 1-dimensional subspace in $\R^4$. There are two nodes of dimension 2, one 3-dimensional node, and one 4-dimensional node in the spine. It is evident from the spine, seen in Figure~\ref{fig:2planesspine}, that it is impossible to move from one of the 2-dimensional maximal nodes to the 3-dimensional maximal node without passing through the 4-dimensional non-maximal node.
Here the cover tree and scaffolding were constructed using eigenthreshold $\tau = 0.5$ and distance threshold $\delta = 0.18$.

\section{Music Structure Visualization}
\label{sec:music}

We turn to the domain of music analysis to test our geometric models on real data.  Specifically, we address the problem of music structure analysis by representing a song as a point cloud in some feature space and using our methods to automatically group distinct musical segments, such as the chorus and verses, into strata.  The problem of automatically recognizing song structure and detecting boundaries between song segments is long standing in the music information retrieval community \cite{paulus2010state}.  Of particular note is recent work in \cite{mcfee2014analyzing} which uses spectral clustering on nearest neighbor graphs of a point cloud in a related feature space to ours, with additional edges added between points adjacent in time.  The eigenvectors corresponding to the lower eigenvalues can then be used to encode membership to different song segments, and taking increasing numbers of eigenvectors allows for a multi-scale notion of structure, which is important, since the notion of a song ``segment" is otherwise ill-posed \cite{mcfee2015_hierarchical_eval}.  Compared to this technique and other previously published techniques, we are less interested in precise segmentation and more interested in the potential for visualization using the spines from our geometric models, but we note that our framework can also naturally handle multi-scale information by varying eigenthresholds or persistence cutoffs in the cover tree construction.

\subsection{Music Feature Space}

Before showing examples of our pipeline on music data, we first provide a high level overview of our scheme for turning music into point clouds which are then plugged into our pipeline.  It is of utmost importance to perform a feature extraction on audio time series, as music data typically consists of very noisy, high frequency data sampled at 22050 or 44100 samples per second which is difficult to analyze in its raw form.
The first group of features we use are referred to as ``timbral features," which are simple shape features on top of the Short-Time Fourier Transform (STFT) frame meant to summarize relative information across the spectrum \cite{tzanetakis2002musical}.  We also use the popular ``Mel-Frequency Cepstral Coefficients" (MFCC) \cite{bogert1963quefrency}, which are perceptually motivated coarse descriptors of the shape of the spectral envelope of an STFT window.  Absolute pitch information is blurred in the above features, so we also use a feature set known as ``chroma" to capture complementary information about notes and chords (see \cite{bartsch2001catch}, \cite{fujishima1999realtime}).

All of the features described so far are typically computed over a very small amount of audio, which we call a ``window," so that each spectrogram frame is nearly stationary (\cite{tzanetakis2002musical} calls this an ``analysis window").  In our application, we typically take this window size to be 2048 samples for audio sampled at 44100hz, or approximately 50 milliseconds.  For 5 timbral features, 12 MFCC coefficients, and 12 chroma features, this gives a total of 29 features per window.  The features in these windows alone are not appropriate for structure modeling, however, because they are simply too short in time, causing rapid variations from one window to the next and the inability to model higher level information.  To make each feature vector more distinct and the resulting point cloud smoother, we aggregate sets of windows into larger time scales called ``blocks" (\cite{tzanetakis2002musical} calls these ``texture windows").  In each block, we take the mean and standard deviation of each feature in the windows contained in that block, for a total of 29*2, adding an additional ``low energy measure" over the block (as recommended in \cite{tzanetakis2002musical}).  This leaves a total of 59 feature dimensions per block.  Since the features are on different scales, we also normalize each dimension by its standard deviation.  In our applications, we typically take the set of all blocks consisting of 150 consecutive windows, spanning roughly 7 seconds per block.  In sum, we transform each song into a 59 dimensional point cloud, where each point represents perceptually motivated statistics in a 7 second window.

\subsection{Spines on Music Data}
\label{sec:MusicSpines}

\begin{figure*}[ht]
(a) \includegraphics[width=0.4\textwidth]{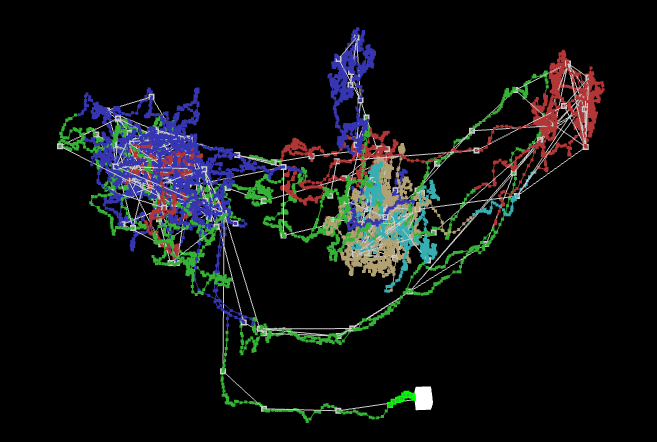} (b) \includegraphics[width=0.45\textwidth]{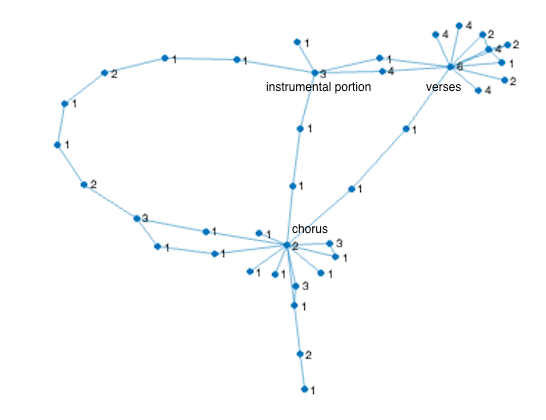} 
\caption{(a) The point cloud corresponding to Michael Jackson's song ``Bad,'' colored by dimension within each node and with the scaffolding edges included.
(b) The spine graph for ``Bad.'' }
\label{fig:MJ}
\end{figure*}

\begin{figure*}
(a) \includegraphics[width=0.35\textwidth]{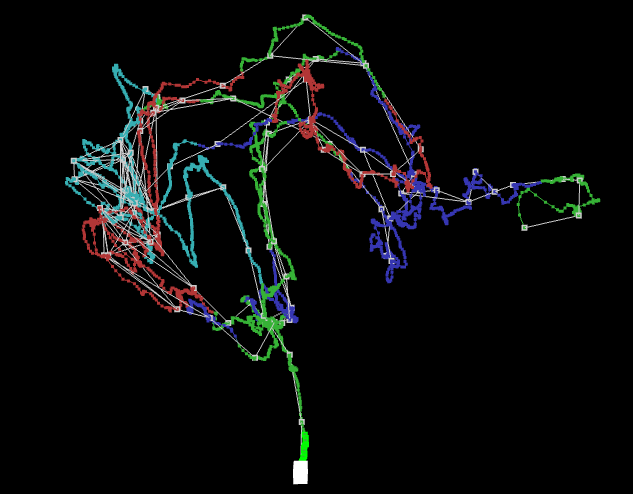} (b) \includegraphics[width=0.45\textwidth]{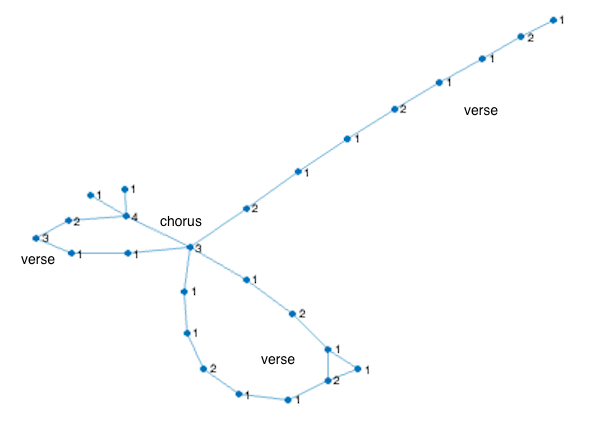}
\caption{The point cloud (a) and spine graph (b) corresponding to The Beatles' song ``Lucy in the Sky with Diamonds.''}
\label{fig:Lucy}
\end{figure*}

We now apply our pipeline on two real songs\footnote{interactive versions of these examples, which allow one to play the song and explore the spine at the same time, can be found at http://www.ctralie.com/Research/GeometricModels} ``Bad" by Michael Jackson (Figure~\ref{fig:MJ} (a) ) and ``Lucy in The Sky with Diamonds" by The Beatles (Figure~\ref{fig:Lucy} (a)). We visualize the point clouds in $\R^3$ by projecting onto the first three principal components (in both cases, approximately $60\%$ of the variance is explained by these components).  The spine graphs (Figure~\ref{fig:MJ} (b) and Figure~\ref{fig:Lucy} (b)) visibly capture high level structural elements of the music.  Areas of greater musical complexity (e.g., verses, chorus) tend to be characterized by higher dimensional strata, whereas transitions between distinct parts of the song are represented by one-dimensional strata.  In ``Bad," an additional distinction is seen between the instrumental only part of the verse and the part with the same vamp (repeating) instrumentals with vocals on top, which are acoustically similar but different enough to be clustered into different strata right next to each other in our scheme.  Overall, our spines provide musically relevant visuals for acoustic clustering and transitions.

\section{Discussion}
\label{sec:discussion}

This paper proposes a flexible and fast new technique for high-dimensional data organization, visualization, and analysis. The most interesting datasets to apply this method to are those sampled from stratified spaces. We have demonstrated the utility of the method with some experiments on synthetic data, and our main application in this work was to music structure analysis. Going forward, we plan to build on our methods to perform classification by genre or by artist, and we could also apply our framework to a time series corpus with a metric between different time series, in addition to applying it to sliding windows of a single time series as we did in Section~\ref{sec:MusicSpines}.

Our framework is quite general and is well-suited for many kinds of data, not just music. For instance, in the domain of natural language processing, we could analyze a collection of documents using some metric between them, with the aim of separating out different topics into different strata.  With a simple bag of words approach, there is also a natural interpretation of the cover tree centers and eigenvectors within each strata: the centers capture the words that make up the main topic/conversation, while the eigenvectors capture groups of words that differentiate documents in the same conversation. In a different application, we could apply our pipeline to group EKG measurements for different types of heart disease patients. Finally, we could use our pipeline to help users efficiently navigate large collections of images or even of 3D shapes (building off of the work of \cite{kim2013understanding}).




We end by outlining some potential conjectures that we plan to pursue in future work.
First, the scaffolding $\Sigma(\X)$ and the spine $S(\X)$ are both representations of the point cloud $\X$. It would be nice to be able to show some sort of stability result relating these to $\Sigma(\Y)$ and/or $S(\Y)$ for some point cloud $\Y$ that was near $\X$ in, say, Hausdorff distance. 
Second and finally, suppose that $(X,<)$ is a partially ordered set \cite{Birkhoff1948}. We recall that the \emph{Haase diagram} of this partial ordering is a directed graph $G = (X,E)$,
where there is an arc from $x$ to $y$ if and only if $x < y$. If $\{S_i\}$ is a stratification of a space $Y$, then there is a partial ordering on the set $\mathcal{S}$ of strata, where if $S_i < S_j$, the local neighborhood of a point in $S_i$ looks like the intersection of $S_j$ with one or more other strata.
In each of the examples we've shown, the spine of our point cloud $\X$ is exactly the Haase diagram of the partial ordering on the stratification of the space $Y$ that $\X$ was sampled from.
Compare, for example, the spine in part (c) of Figure~\ref{fig:biggraph} with the space underlying the point cloud in part (a); arrows are not drawn, but imagine that the lower index always points to the higher. 
Of course, parameters had to be chosen to make this true. We conjecture that this should always work, assuming some appropriate sampling conditions.


\section*{Acknowledgments}

The first and third authors were partially supported by NSF award BIGDATA 1444791, and the first was also partially supported by NSF award WBSE 3331753. 
The fourth author was partially supported by the Research Training Grant NSF-DMS award 1045133 and an NSF Graduate Fellowship. 
The second author thanks the Department of Mathematics at Duke University for hosting her during the fall of $2015$.

\clearpage

\bibliography{GeometricModels}{}
\bibliographystyle{plain}

\end{document}